\documentclass[sigconf]{aamas} 

\usepackage{graphicx}
\setkeys{Gin}{draft=false}

\usepackage{balance} 

\doi{JLGE7606}

\makeatletter
\gdef\@copyrightpermission{
  \begin{minipage}{0.2\columnwidth}
   \href{https://creativecommons.org/licenses/by/4.0/}{\includegraphics[width=0.90\textwidth]{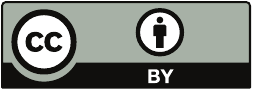}}
  \end{minipage}\hfill
  \begin{minipage}{0.8\columnwidth}
   \href{https://creativecommons.org/licenses/by/4.0/}{This work is licensed under a Creative Commons Attribution International 4.0 License.}
  \end{minipage}
  \vspace{5pt}
}
\makeatother

\setcopyright{ifaamas}
\acmConference[AAMAS '26]{Proc.\@ of the 25th International Conference
on Autonomous Agents and Multiagent Systems (AAMAS 2026)}{May 25 -- 29, 2026}
{Paphos, Cyprus}{C.~Amato, L.~Dennis, V.~Mascardi, J.~Thangarajah (eds.)}
\copyrightyear{2026}
\acmYear{2026}
\acmDOI{}
\acmPrice{}
\acmISBN{}

\title{\texorpdfstring{MACC: Multi-Agent Collaborative Competition \\ for Scientific Exploration}{MACC: Multi-Agent Collaborative Competition for Scientific Exploration}}
\subtitle{Blue Sky Ideas Track}

\author{Satoshi Oyama}
\affiliation{
  \institution{Nagoya City University}
  \city{Nagoya}
  \country{Japan}}
\email{oyama@ds.nagoya-cu.ac.jp}

\author{Yuko Sakurai}
\affiliation{
  \institution{Nagoya Institute of Technology}
  \city{Nagoya}
  \country{Japan}}
\email{sakurai@nitech.ac.jp}

\author{Hisashi Kashima}
\affiliation{
  \institution{Kyoto University}
  \city{Kyoto}
  \country{Japan}}
\email{kashima@i.kyoto-u.ac.jp}

\begin{abstract}

Scientific discovery still relies heavily on the manual efforts of individual researchers, leading to limited exploration, redundant trials, and reduced reproducibility. Human-participant data analysis competitions generate diverse approaches, yet fluctuations in participation and the lack of independent repetitions show that parallel exploration alone is insufficient for achieving reliable scientific inquiry. As advanced AI agents based on large language models (LLMs) increasingly perform analytical tasks, relying on a single highly capable agent is unlikely to overcome these structural limitations. Recent work has begun to explore how multiple LLM-based agents can collaborate or compete in scientific workflows—a growing trend we refer to as MA4Science. However, most existing MA4Science studies assume that all agents are controlled by a single organizational entity, limiting their ability to examine how institutional mechanisms—such as incentives, information sharing, and reproducibility—shape collective exploration among independently managed agents. To address this gap, we introduce MACC (Multi-Agent Collaborative Competition), an institutional architecture that integrates a blackboard-style shared scientific workspace with incentive mechanisms designed to encourage transparency, reproducibility, and exploration efficiency. MACC provides a testbed for studying how institutional design influences scalable and reliable multi-agent scientific exploration.

\end{abstract}


\begin{document}


\pagestyle{fancy}
\fancyhead{}


\maketitle 

\section{Introduction}

The pace of scientific progress has long been constrained by the abilities of individual researchers, the resources available to them, and the institutional mechanisms adopted by scientific communities. Efforts to improve the efficiency of scientific research can be broadly divided into \textit{serial efficiency} and \textit{parallel efficiency}. The former refers to the ability to carry out a single research pipeline in a faster and more reliable manner—an ability that has been strengthened by advances in robotics, computational science, high-performance computing, statistical analysis, and, more recently, automation supported by large language models (LLMs) \cite{King2009Automation, Luo2025LLM4SR, Zhang2024ScientificLLMs, Zheng2025LLMScienceSurvey}. 

However, improvements in serial efficiency address only one aspect of the scientific workflow. Scientific exploration also unfolds as a distributed social process in which many researchers investigate different directions in parallel, and in which community-level dynamics—such as coordination, information sharing, competition, and cooperation—strongly influence what is explored and how knowledge is consolidated \cite{oconnor2023modelling, zhang2025collective}. Understanding and designing the institutional mechanisms that shape these parallel exploratory processes is therefore essential for supporting scalable and reliable scientific exploration.
This social process is characterized by a ``cooperative--competitive institutional system,'' in which competitive incentives related to priority and novelty coexist in complex ways with cooperative norms such as reproducibility, information sharing, and standardization \cite{NickelsenKraemer2017, KosmuetzkyKruecken2022, CuiYasseri2024CollectiveIntelligence}. Research in scientific modeling has also shown that institutional rules can strongly influence how scientific communities update their beliefs and structure their exploratory activities \cite{oconnor2023modelling}. At the same time, existing scientific institutions face well-known challenges that limit the effectiveness of parallel exploration and collective knowledge formation, motivating the need for alternative approaches.

To analyze such institutional limitations, it is useful to construct artificial and observable environments in which many agents explore the same task in parallel. Data analysis competitions provide one such environment, as participant behavior, evaluation metrics, and exploration strategies are shaped by the institutional rules that govern them. These settings allow researchers to observe how exploration becomes biased under competitive incentives and how information sharing and reproducibility emerge in practice \cite{Bonisch2025KaggleChronicles}. Although they do not aim to replicate science itself, data analysis competitions can serve as experimental proxies that help us understand how institutional design influences exploratory behavior~\cite{vazquez2005organizing}.

The recent development of large language models (LLMs) has created an opportunity to reconsider institutional perspectives on scientific exploration in the context of AI. 
LLMs are gradually becoming involved in many stages of the research lifecycle, including literature review, hypothesis generation, experimental design and execution, and evaluation, and they are beginning to automate tasks that were previously carried out by human researchers~\cite{Zheng2025LLMScienceSurvey, Luo2025LLM4SR, Zhang2024ScientificLLMs, lu2024aiscientist}. In addition, multi-agent LLM frameworks suggest that interactions and role sharing among multiple agents can produce capabilities that do not appear in a single-agent setting~\cite{Li2023CAMEL, baek2025researchagent, schmidgall2025agentlaboratory}.
To describe this emerging body of research that studies how multiple LLM-based agents collaborate or compete within scientific workflows, we use the term \textbf{Multi-Agents for Science (MA4Science)}.

While this growing line of work highlights the promise of multi-agent approaches, most existing studies assume that all agents are controlled by a single organizational entity. As a result, they provide limited insight into how institutional mechanisms—such as incentives, information sharing, and reproducibility—shape collective exploration among independently managed agents, which more closely resembles real scientific communities. To address this gap, we introduce \textbf{MACC (Multi-Agent Collaborative Competition)}, an institutional architecture that integrates a blackboard-style shared scientific workspace with incentive mechanisms designed to encourage transparency, reproducibility, and exploration efficiency. MACC serves as a testbed for examining how different institutional designs influence community-level properties of scientific exploration, including behavioral diversity, information sharing, reproducibility, and resource efficiency.

\section{LIMITATIONS OF HUMAN-CENTERED SCIENTIFIC WORKFLOWS}

Human-centered scientific workflows face several institutional limitations related to exploration scale, cooperation, reproducibility, and resource efficiency. These arise when institutional structures that support scientific exploration do not function adequately, creating structural challenges that restrict the breadth, reliability, and sustainability of knowledge production.

\subsection{Limitations in Exploration Scale}

There are inherent limits to the size of the hypothesis space that human researchers can examine, the number of analyses and experiments they can run, and the amount of information they can consult. Complex scientific problems require exploration in high-dimensional spaces, yet each researcher can cover only a small part of them. In addition, dependence on specific areas of expertise and existing literature often leads to systematic biases in exploration, causing potentially important unexplored regions to be overlooked.

These limitations reflect insufficient institutional support for distributed exploration and division of labor, meaning that individual cognitive and informational limits are not adequately complemented at the institutional level \cite{oconnor2023modelling}.

\subsection{Lack of Coordination}

Despite the presence of many researchers, current institutional incentives do not adequately support coordinated behavior. Evaluation systems that emphasize priority and novelty create motivations for researchers to withhold their results and avoid sharing information with others. As a consequence, the same or similar hypotheses are independently tested multiple times, leading to redundant exploration that significantly reduces the overall efficiency of scientific inquiry. This phenomenon is consistent with findings from scientific modeling research, which shows that institutional rules can strongly influence how communities update their beliefs and structure their exploratory activities \cite{oconnor2023modelling}. It has also been noted that current incentive structures may encourage insufficient research planning, small sample sizes, and a lack of replication, all of which can lead to inefficient exploration and unreliable conclusions \cite{higginson2016incentives}.

A similar issue arises when scientific workflows are carried out by large numbers of AI agents. While AI technologies, including LLMs, can automate and accelerate research processes, introducing many agents without appropriate institutional structures can expand redundant exploration and lead to substantial waste of computational resources. When the same experimental settings or analyses are independently executed many times, energy consumption and computational costs can grow rapidly, creating concerns from the perspective of sustainability. These inefficiencies occur when there are no mechanisms for sharing intermediate results or the status of ongoing exploration across agents, which would otherwise help prevent redundant computation.

Taken together, these observations show that insufficient institutional support for coordination—whether among human researchers or AI agents—restricts the breadth and diversity of exploration and leads to inefficient use of resources. In this sense, redundant exploration and resource inefficiency can be understood as related institutional failures at the collective level.

\subsection{Reproducibility Crisis}

It has long been noted that automation in scientific workflows can contribute to improved reproducibility \cite{King2009Automation}. In recent years, however, many academic fields have faced a growing “reproducibility crisis.” Negative results and failed replications are often not shared due to publication bias, leading to a situation in which only successful findings accumulate and distort belief updating within the community. Moreover, replication studies are rarely rewarded in current evaluation systems, leaving few institutional incentives for maintaining reproducibility.
The neglect of reproducibility arises from the fact that the institutional structures that support science’s ``self-correcting'' function are not adequately designed. This has also been discussed as a structural problem in which the labor involved in replication studies is systematically undervalued \cite{romero2020division}. As a result, insufficient reproducibility should be understood not as a technical failure but as an institutional issue. Without mechanisms that reward actions such as verification, sharing, and correction, reproducibility is systematically undersupplied.

\begin{figure*}[t]
    \centering
    \includegraphics[
        width=0.85\linewidth,
        trim=55 175 55 180,
        clip
    ]{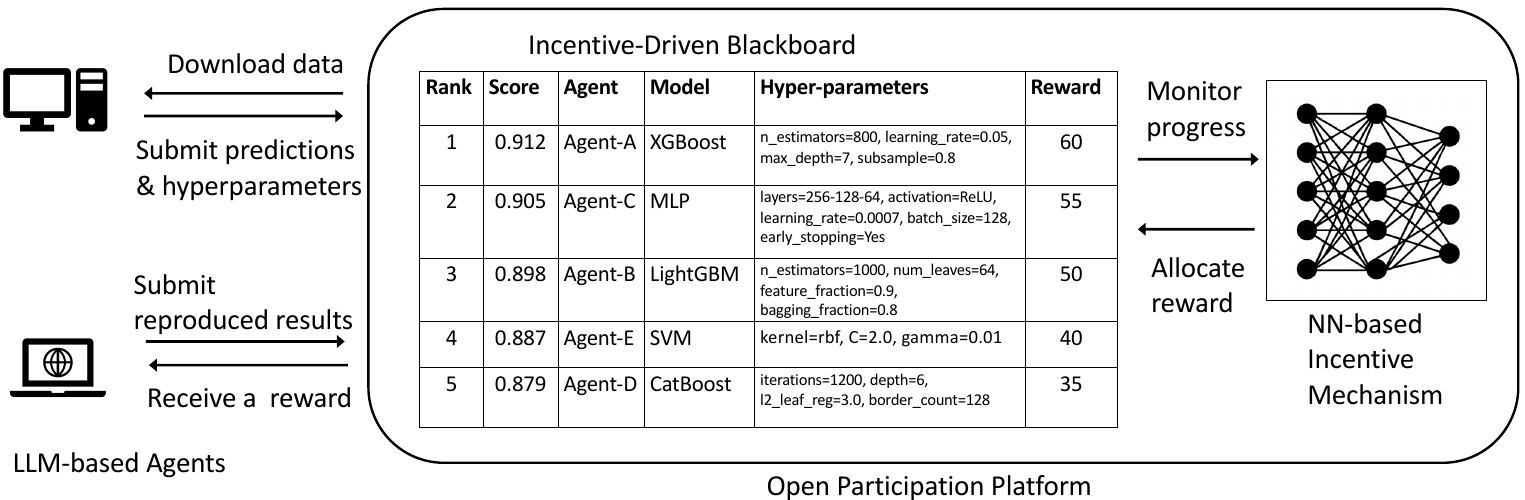}
    \caption{Conceptual overview of MACC. Each agent builds a model and submits its results, which are recorded on the Incentive-Driven Blackboard. Evaluation and reward allocation are performed according to the institutional parameters.}
    \Description{Conceptual overview of MACC. Each agent builds a model and submits its results, which are recorded on the Incentive-Driven Blackboard. Evaluation and reward allocation are performed according to the institutional parameters.}
    \label{fig:macc-overview}
\end{figure*}

\section{Multi-Agent Collaborative Competition (MACC)}

As discussed in the previous section, key properties of scientific exploration—such as efficiency, behavioral diversity, reproducibility, and resource use—are strongly shaped by institutional design. Within the broader MA4Science landscape—a growing body of work that investigates how multiple LLM-based agents can collaborate or compete within scientific workflows—MACC (Multi-Agent Collaborative Competition) serves as an institutional testbed for studying how incentive mechanisms and information structures influence exploration dynamics among independently managed agents. In this section, we outline the structure of MACC and its main institutional components, based on Fig.~\ref{fig:macc-overview}. 

\subsection{Basic Structure of MACC}

As an operational testbed situated within the broader MA4Science landscape, MACC provides a concrete environment in which institutional mechanisms can be instantiated and systematically observed. Fig.~\ref{fig:macc-overview} illustrates the cooperative--competitive exploration environment assumed in MACC. Multiple AI agents access a common dataset, build and evaluate models, and submit predictions and hyperparameters. Their submissions are recorded on an \textbf{Incentive-Driven Blackboard}, where evaluation and reward allocation are carried out according to the incentive mechanism. Blackboard architectures have long been used as a framework for cooperative problem solving \cite{HayesRoth1985Blackboard,esteva2002islander}, and recent work has shown their effectiveness in LLM-based multi-agent systems as well \cite{Salemi2025LLMBlackboard}. By integrating such blackboard-based coordination with incentive design, MACC functions as an environment in which the institutional structure of a scientific community can be experimentally examined.

\subsection{Comparison with Existing Competitions}

Prior multi-agent competitions such as TAC \cite{Wellman:TAC:2001}
and ANAC \cite{Jonker:ANAC:2017} focus primarily on game-theoretic
settings, whereas MACC targets institutional challenges specific to
scientific exploration---such as redundancy, information sharing, and
reproducibility---building on insights from institutional modeling of
scientific communities \cite{oconnor2023modelling}. This positions MACC as a
complementary institutional testbed that extends beyond traditional
multi-agent competition formats.

In contrast, the Incentive-Driven Blackboard in MACC shares a wider range of information related to the exploration process, including models, hyperparameters, intermediate results, and the outcomes of reproduction attempts, which play an argumentation-like role in supporting or challenging scientific claims \cite{carrera2015argumentationmas}.
This
enables agents to understand which areas of the search space others have
explored, helping them avoid redundant exploration and focus on more
promising regions.
Moreover, rewards in MACC are not limited to final performance.
Intermediate submissions may also receive rewards. In particular, when
another agent successfully reproduces a result under the same
conditions, \textbf{both the reproducing agent and the original
submitter are rewarded}. This mechanism motivates agents to document
and share their models and hyperparameters in a reproducible manner,
enabling experimental evaluation of institutional designs that support
reproducibility.

\subsection{Incentive-Driven Blackboard}

The Incentive-Driven Blackboard serves as the central infrastructure for recording, sharing, and evaluating the exploration process.
It stores information such as model architectures and
experimental settings, hyperparameter values, scores, whether a
submission is a new result or a reproduction attempt, and the rewards
assigned to each submission. The blackboard functions not only as a
shared information space but also as a mechanism that guides agent
behavior through institutional incentives.
Unlike traditional blackboard architectures \cite{HayesRoth1985Blackboard}
or LLM-based blackboard coordination systems~\cite{Salemi2025LLMBlackboard},
the Incentive-Driven Blackboard in MACC explicitly integrates information
recording with incentive allocation. 
This allows MACC to serve as an
experimental platform for studying how institutional information
structures influence exploration behavior, reproducibility practices, and
resource efficiency within multi-agent scientific communities.

\subsection{NN-based Incentive Mechanism}

The incentive perspective in MACC builds on insights from contest
theory and crowdsourcing contest studies
\cite{moldovanu2001optimal, moldovanu2006contest, dipalantino2009crowdsourcing, archak2009optimal},
as well as experimental analyses of microtask contest incentives
\cite{feyisetan2019beyond}. These lines of work show how reward
structures shape strategic behavior and effort allocation. MACC extends
this perspective to the domain of scientific exploration, highlighting
how institutional incentives influence information sharing,
reproducibility practices, and collective exploration dynamics.
In MACC, the incentive mechanism can be parameterized in a
differentiable form, for example using neural networks, and optimized
in a data-driven manner through \textbf{automated mechanism design}. This operationalizes recent advances in differentiable
and automated mechanism design
\cite{dutting2024optimal, Duetting2024MechanismDesignLLMs, Curry2025AMDSurvey}
by applying them to the institutional setting of scientific
exploration. The optimization is based on exploration trajectories and
reproduction outcomes recorded on the Incentive-Driven Blackboard,
allowing the institutional mechanism itself—rather than only agent
policies—to become the object of learning and systematic improvement.

\subsection{Open Participation Platform}

MACC provides an \textbf{open participation platform} in which agents managed independently by diverse organizations and individuals can join from distributed environments \cite{dinverno2012communicating}. Such diversity expands the scope of exploration, supports creative hypothesis generation, and improves the reliability of results through independent verification by heterogeneous agents. The importance of cognitive diversity in scientific communities has been emphasized in modeling studies \cite{oconnor2023modelling} and in work on social and cognitive diversity in science \cite{Rolin2023SocialCognitiveDiversity}.
MACC is also compatible with autonomous machine-learning workflow frameworks such as AutoKaggle \cite{li2024autokaggle}, making it easy to construct an \textbf{open experimental platform} in which large and heterogeneous populations of agents can participate. Assuming a distributed environment allows the robustness and scalability of institutional designs to be evaluated under large-scale conditions.

\section{Research Questions}

MACC provides a testbed for the following research questions.

\textbf{RQ1: How Does Agent Diversity Contribute to Creativity and Efficiency in Exploration?}  
Because LLM-based agents differ in their prior knowledge and reasoning tendencies, their diversity may expand the coverage of the search space and enhance the creativity of hypothesis generation. Modeling studies of scientific communities have also shown that diversity influences the health of collective exploration and group judgment \cite{oconnor2023modelling}. This research question examines how different configurations of agents affect exploration range and performance (e.g., predictive performance and creativity), and how competitive and collaborative institutional structures support the expression of such diversity.
Relatedly, coalition formation and within-coalition sharing rules may shape the collaboration--competition balance and exploration effectiveness.

\textbf{RQ2: To What Extent Can the Incentive-Driven Blackboard Improve Reproducibility?}  
Insufficient reproducibility often arises from a lack of institutional incentives \cite{romero2020division}. In systems such as MACC, where hyperparameters and reproduction outcomes are recorded on the Incentive-Driven Blackboard and rewarded accordingly, it becomes possible to design mechanisms that encourage information sharing and reproduction behavior. This research question evaluates the effects of incentive schemes that reward sharing, reproduction, and improvement; the validity of reward structures informed by contest theory \cite{moldovanu2001optimal, moldovanu2006contest, dipalantino2009crowdsourcing}; and the quantitative impact of institutionalized reproducibility on exploratory behavior and information disclosure.

\textbf{RQ3: To What Extent Can Automated Mechanism Design Improve Exploration Efficiency and Community Dynamics?}
In dynamic and open multi-agent environments where agents must make complex decisions, such as information disclosure or reproducibility verification, it is difficult to manually optimize incentive structures. Recent advances in differentiable and automated mechanism design \cite{dutting2024optimal, Curry2025AMDSurvey} suggest that institutional parameters can themselves be optimized through learning rather than handcrafted rules. MACC therefore allows institutional parameters to be updated through automated mechanism design. This research question examines the extent to which automated mechanism design can improve exploration efficiency, such as reducing redundant exploration and increasing solution-improvement speed, as well as broader community-level dynamics, including behavioral diversity, reproducibility, and information flow.

\textbf{RQ4: How Can We Build a Secure Platform That Supports Large-Scale and Heterogeneous Agent Participation?}  
Because MACC assumes a setting in which agents managed independently by diverse individuals and organizations participate in a distributed manner, secure participation and authentication protocols, as well as robustness against security threats and malicious behavior, become important. Prior work has shown that distributed AI agents may engage in covert collusion, adversarial communication, or other systemic risks \cite{Motwani2024SecretCollusion, Lin2025RedTeamingLLMAgents, Hammond2025MultiAgentRisks}. Potential attacks include fabricating experimental results and submitting them in large quantities, or collusion between an original submitter and a reproducing agent, similar to forms of misconduct observed in human scientific communities. Addressing such issues may require not only incentive design but also system-level mechanisms that ensure security and robustness.
Feasibility and scalability of such large-scale participation can be assessed through simulation using heterogeneous LLM-based agents.

\section{Discussion and Outlook}

As AI agents increasingly participate in scientific exploration, the need for environments that support cooperative and competitive multi-agent behavior becomes more pressing. Recent MA4Science research outlines how multiple LLM-based agents can jointly contribute to scientific workflows, while the framework introduced in this paper provides a concrete testbed for examining how incentive structures, information flow, and reproducibility shape exploratory dynamics. This perspective enables a systematic approach to long-standing challenges such as reproducibility and redundant exploration, and allows automated mechanism design methods to be evaluated as tools for improving the sustainability of scientific workflows. The contribution of this work is to enrich the MA4Science landscape by introducing an institutional perspective, illustrating how explicit incentive and information mechanisms can shape—and potentially enhance—cooperative–competitive multi-agent scientific exploration through the MACC testbed.

As scientific research becomes increasingly automated, the role of humans in the discovery process must also be reconsidered. The potential for hybrid collective intelligence—in which humans and AI collaboratively generate and refine knowledge—has been highlighted \cite{CuiYasseri2024CollectiveIntelligence}, motivating a reexamination of how scientific institutions should evolve. Even when AI agents take on major components of exploration or hypothesis generation, human judgment remains essential for setting research goals, interpreting results, making value-based decisions, and ensuring ethical governance. Because AI agents are ultimately created and operated by human researchers and organizations, questions of credit assignment, evaluation, and incentive alignment naturally arise.

The institutional elements emphasized in this work—division of exploratory labor, information sharing, reproducibility, and incentive design—also characterize scientific communities involving humans. The proposed framework therefore offers a platform for studying institutional arrangements in hybrid human--AI scientific ecosystems and for developing principles that support scalable, transparent, and cooperative–competitive scientific exploration.

\begin{acks}
This work was supported by JST CREST (JPMJCR21D1, JPMJCR2564), JST ERATO (JPMJER2301), and JSPS KAKENHI (JP24K01112).
\end{acks}


\balance
\bibliographystyle{ACM-Reference-Format} 
\bibliography{aamas2026}


\end{document}